\documentclass[a4paper,english]{article}
\usepackage[T1]{fontenc}
\usepackage[latin1]{inputenc}
\usepackage{amstext}
\usepackage{amssymb}



\usepackage{babel}
\begin{document}

\title{Giant dipole oscillations and ionization of heavy atoms by intense
electromagnetic fields}

\author{{\normalsize{M. Apostol }}\\
{\normalsize{Department of Theoretical Physics, Institute of Atomic
Physics,}}\\
{\normalsize{ Magurele-Bucharest MG-6, POBox MG-35, Romania}}\\
{\normalsize{ email:apoma@theory.nipne.ro}}}

\date{{}}

\maketitle
\relax
\begin{abstract}
The ''empirical'' binding energy $-16Z^{7/3}eV$ of heavy atoms
(atomic number $Z\gg1$) is computed by a linearized version of the
Thomas-Fermi model, including a Hartree-type correction. The computations
are carried out by means of a variational approach. Exchange energy
and corrections to the exchange energy are also estimated. This is
an updated result. It is shown that giant dipole oscillations of the
electrons may be induced in heavy atoms by external electromagnetic
fields in the range of moderate $X$-rays, which, in intense fields,
may lead to ionization. There are examined anharmonicities in the
giant dipole oscillations, which lead to frequeny shifts and high-order
harmonics. Transitions to excited states and ionization of \textquotedbl{}peripheral\textquotedbl{}
electrons are also investigated in the quasi-classical approximation
for heavy atoms.
\end{abstract}
\relax



\noindent \textbf{Introduction. Thomas-Fermi model.} The Thomas-Fermi
model is based on the quasi-classical description and the statistical
character of the electron single-particle states in heavy atoms, \textit{i.e.}
in atoms with large atomic numbers $Z$ ($Z\gg1$).\cite{key-1}-\cite{key-4}
The electrons are assumed to form an inhomogeneous, dense gas of fermions,
slightly perturbed by the nuclear charge $Ze$, where $-e$ is the
electron charge. The singular character of the nuclear Coulomb potential
$Ze/r$ at the origin is compensated by its relatively small range
around the origin, which is left by the electron screening; this particularity
justifies the perturbation character of the Coulomb potential. 

\noindent Usually, the Fermi wavevector $k_{F}$ is determined by
a self-consistent field potential $\varphi$, according to 
\begin{equation}
\hbar^{2}k_{F}^{2}/2m-e\varphi=0\,\,,\,\, k_{F}=(2me/\hbar^{2})^{1/2}\varphi^{1/2}\,\,\,,\label{1}
\end{equation}
 where $\hbar$ is Planck's constant and $m$ is the electron mass.
Equation (\ref{1}) expresses the energy conservation, with zero total
energy for neutral atoms. The electric potential $\varphi$ is determined
by Poisson's equation 
\begin{equation}
\Delta\varphi=-4\pi Ze\delta({\bf r})+4\pi ne\,\,\,,\label{2}
\end{equation}
 where the electron density is given by $n=k_{F}^{3}/3\pi^{2}=(1/3\pi^{2})(2me/\hbar^{2})^{3/2}\varphi^{3/2}$,
as for a Fermi gas; equation (\ref{2}) becomes \textquotedbl{}the
$3/2$\textquotedbl{}-equation, which is solved numerically with the
boundary condition $\varphi r\rightarrow0$ for $r\rightarrow\infty$
(and $\varphi=Ze/r$ for $r\rightarrow0$).\cite{key-5} The atomic
binding energy, as computed by means of this theory,\cite{key-4},\cite{key-6,key-7}
is given by $E\simeq-20.8Z^{7/3}{\rm eV}$, which is the exact result
in the limit $Z\rightarrow\infty$.\cite{key-8} This means that convincing
arguments have been presented\cite{key-8} that the Schrodinger equation
for $Z$ electrons in the coulombian field of the neutralizing atomic
nucleus gives the ground-state energy $E\simeq-20.8Z^{7/3}{\rm eV}$
in the limit $Z\rightarrow\infty$. Corrections have been brought
to this result, the binding energy being represented as an asymptotic
series in powers of $Z^{-1/3}$; this series includes, beside the
leading term $-20.8Z^{7/3}{\rm eV}$, the so-called ''boundary correction''
$13.6Z^{2}$eV,\cite{key-9,key-10} the exchange contribution $-5.98Z^{5/3}$eV
(or $-7.3Z^{5/3}$eV),\cite{key-11} etc; in addition, the relativistic
effects must also be considered for large $Z$.

\noindent In principle, such an asymptotic series should reproduce
satisfactorily the experimental atomic binding energies; this would
be an illustration of the \textquotedbl{}unreasonable utility of asymptotic
estimates''.\cite{key-12} In fact, the empirical binding energy
of heavy atoms is well represented by $E\simeq-16Z^{7/3}{\rm eV}$
(see, for instance Ref. \cite{key-9} and references quoted therein),
which differs appreciably from the leading term of the asymptotic
series.

\noindent In order to improve the results various computations have
been worked out, including higher-order corrections to the quasi-classical
approximation, self-consistent Hartree, or Hartree-Fock equations,
as well as density-functional models.\cite{key-13}-\cite{key-15}
At the same time, the Thomas-Fermi model revealed another drawback:
it does not bind atoms in molecules.\cite{key-16,key-17} 

\noindent We present here a different approach to the problem, which
provides a more direct access to the $E\simeq-16Z^{7/3}{\rm eV}$-representation
of the ''empirical'' binding energies of the atoms, and may throw
additional light upon the nature of the Thomas-Fermi model and the
quasi-classical description. The method employed here is a variational
treatment of a linearized version of the Thomas-Fermi model, as based
on the quasi-classical description. In particular, it binds atoms
in large clusters.

\noindent \textbf{Linearized Thomas-Fermi model.\cite{key-18}} According
to the prescriptions of the quasi-classical approximation, equation
(\ref{1}) is valid as long as the potential $\varphi$ varies slowly
in space; consequently, the Fermi wavevector $k_{F}$ has also a slow
spatial variation, so one may linearize equation (\ref{1}) by substituting
$2\overline{k}_{F}k_{F}$ for $k_{F}^{2}$, where $\overline{k}_{F}$
is viewed as a variational parameter, assumed to be constant in space,
the whole spatial dependence being transferred upon the new variable
$k_{F}$; this substitution is justified for those spatial regions
where $\overline{k}_{F}$ and $k_{F}$ are comparable in magnitude,
and one can see easily that this is so for a moderate range of intermediate
distances; it is precisely this range over which the most part of
the electrons are localized in heavy atoms, so that one may expect
to get a reasonable description by employing this linearization procedure.
Alternatively, we may consider the interaction as a small perturbation
and write equation (\ref{1}) as $\frac{\hbar^{2}}{m}k_{F}\delta k_{F}-e\delta\varphi=0\,\,;$
here we replace $k_{F}$ by $\overline{k}_{F}$, $\delta k_{F}$ by
$k_{F}$ and $\delta\varphi$ by $\varphi$. As remarked before, it
is worth noting that although the Coulomb potential is singular at
the origin, it extends over a small region around the origin, due
to the screening, which makes its effects suitable to be treated as
a small perturbation. On the other side, the original dependence of
the electron density on the $3/2$-power of the potential $\varphi$,
$n\sim\varphi^{3/2}\sim1/r^{3/2}$, overestimates the density in the
zone of the abrupt variation of the potential $\varphi$, \emph{i.e.}
near the nucleus, where the potential goes like $\varphi\sim Ze/r$,
in comparison with the linearized version $n\sim\varphi\sim1/r$,
which is contrary to the requirements of the quasi-classical approximation.
We get $k_{F}=(me/\hbar^{2}\overline{k}_{F})\varphi$ for the linearized
version of equation (\ref{1}). A similar linearization for the electron
density $n=k_{F}^{3}/3\pi^{2}\rightarrow n=\overline{k}_{F}^{2}k_{F}/\pi^{2}$
leads to $n=(me\overline{k}_{F}/\pi^{2}\hbar^{2})\varphi=(q^{2}/4\pi e)\varphi$,
where the Thomas-Fermi screening wavevector $q$ has been introduced
through $q^{2}=4me^{2}\overline{k}_{F}/\pi\hbar^{2}=4\overline{k}_{F}/\pi a_{H}$
(here $a_{H}=\hbar^{2}/me^{2}\simeq0.53\textrm{\AA}$ is the Bohr
radius). Now, Poisson's equation (\ref{2}) has the well-known solution
$\varphi=(Ze/r)e^{-qr}$, \textit{i.e.} the screened Coulomb potential,
as expected. One can see that this potential falls abruptly to zero
at large distance, where the quasi-classical description does not
appply (as the wavelengths increase indefinitely there), varies slowly
over intermediate distances, as expected, and has an abrupt variation
over short distances, \textit{i.e.} near the atomic nucleus; in the
small region around the nucleus the computations will be corrected,
as required by the quantum behaviour of the electrons in this region.
For the moment, however, we proceed further on, by computing the total
energy.

\noindent By using the same linearization procedure the kinetic energy
$E_{kin}=V\hbar^{2}k_{F}^{5}/10\pi^{2}m$ of the electron gas enclosed
in a volume $V$ is replaced by 
\begin{equation}
E_{kin}=(\hbar^{2}\overline{k}_{F}^{4}/2\pi^{2}m)\int d{\bf r}\cdot k_{F}=\frac{\pi ea_{H}^{3}}{128}q^{6}\int d{\bf r}\cdot\varphi\,\,\,,\label{3}
\end{equation}
 which yields $E_{kin}=\pi^{2}a_{H}^{3}Ze^{2}q^{4}/32$. The potential
energy is given by 
\begin{equation}
\begin{array}{c}
E_{pot}=\int d\mathbf{r}(\rho_{e}\varphi-\frac{1}{2}\rho_{e}\varphi_{e})=\frac{1}{2}\int d\mathbf{r}(\rho_{e}\varphi+\rho_{e}\varphi_{c})=\\
\\
=-\frac{e}{2}\int d\mathbf{r}n(\varphi+\varphi_{c})=-\frac{q^{2}}{8\pi}\int d\mathbf{r}(\varphi^{2}+\varphi\varphi_{c})\,\,\,,
\end{array}\label{4}
\end{equation}
 where $\rho_{e}=-en$ is the density of the electronic charge, $\varphi_{e}=\varphi-\varphi_{c}$
is the electric potential produced by the electrons, and $\varphi_{c}=Ze/r$
is the Coulomb potential of the atomic nucleus. The computations are
straightforward, and one obtains $E_{pot}=-3Z^{2}e^{2}q/4$. Therefore,
the total energy reads 
\begin{equation}
E=E_{kin}+E_{pot}=\frac{\pi^{2}a_{H}^{3}}{32}Ze^{2}q^{4}-\frac{3}{4}Z^{2}e^{2}q\,\,\,,\label{5}
\end{equation}
 which reaches the minimum value 
\begin{equation}
E=-\frac{9\cdot6^{1/3}}{16\pi^{2/3}}Z^{7/3}\frac{e^{2}}{a_{H}}=-0.42Z^{7/3}\frac{e^{2}}{a_{H}}=-11.4Z^{7/3}eV\label{6}
\end{equation}
 for the optimal value 
\begin{equation}
q=(6/\pi^{2})^{1/3}\frac{Z^{1/3}}{a_{H}}=0.85\frac{Z^{1/3}}{a_{H}}\label{7}
\end{equation}
of the variational parameter $q$; we note the occurrence of the atomic
unit for energy $e^{2}/a_{H}\simeq27.2$eV (another useful formula
is $E=-(9/16)Z^{2}e^{2}q$, where $q$ is given by equation (\ref{7})).
(It is woth noting that the potential computed by solving numerically
the $3/2$-equation goes, approximately, like $\varphi\simeq(Ze/r)e^{-2qr}$
for $r\simeq0$, where $q$ is given by equation (\ref{7}) (see,
for instance, Ref. \cite{key-5}); that means that it is more abrupt
near the nucleus than the potential given by the linearized version,
which is a consequence of the overstimation of the electron density
in the vicinity of the nucleus. This leads to an enhanced binding
energy ($-20.8Z^{7/3}{\rm eV}$)). One can see that the radial density
of electrons $\sim r^{2}n$, as given by $n=(q^{2}/4\pi e)\varphi$,
has a maximum value for $R\sim1/q\sim Z^{-1/3}a_{H}$, which may be
taken as the ''radius'' of the electronic charge (while the ''radius''
of the atom is of the order of $a_{H}$); thus, one can see again
that the quasi-classical description for large $Z$ is justified;
indeed, the quasi-classical description holds for distances longer
than the radius $a_{H}/Z$ of the first Bohr orbit and shorter than
the Bohr radius $a_{H}$, and the electronic ''radius'' $R\sim Z^{-1/3}a_{H}$
is such that the inequailities $a_{H}/Z\ll R\sim a_{H}/Z^{1/3}\ll a_{H}$
are satisfied for large $Z$; the most part of the electrons are localized
around $R$, which justifies the statistical character of the Thomas-Fermi
model for large $Z$. The linearization of the basic equations of
the quasi-classical description, together with the variational approach,
as well as the approximate character of the quasi-classical description
in general, which alters the distinction between the exact kinetic
and potential energies, lead to the breakdown of the virial theorem;
indeed, one can check easily that $E_{kin}=-(1/4)E_{pot}$, instead
of $E_{kin}=-(1/2)E_{pot}$, as required by the virial theorem; this
is not a major drawback, as it is well-known that approximate calculations
may give wrong values for both the ''kinetic'' and ''potential''
energies and still the total energy be quite close to the exact one;\cite{key-19}
this is due to the variational treatment employed here.

\noindent \textbf{Quantum correction.} According to equation (\ref{1})
and the Thomas-Fermi model, the electronic states are described by
quasi-plane waves everywhere in space, their wavevector depending
weakly on position; they correspond to the electron motion in a slowly-varying
potential, vanishing at large distances; the screened Coulomb potential
$\varphi$ is consistent with this assumption, except for short distances
where it has a sudden variation; the electron single-particle energies
must therefore be corrected for this additional potential energy,
corresponding to the electron motion close to the atomic nucleus;
the correction is carried out to the first order of the perturbation
theory, by estimating the average of the potential energy $-e\varphi$
over plane waves confined to a small spherical region of radius $R$
around the nucleus; the radius $R$ must be regarded as a variational
parameter, and the correction to the energy will be minimized with
respect to $R$; doing so, we obtain an additional energy 
\begin{equation}
-\frac{e}{v}\int_{v}d{\bf r}\cdot\varphi=-3Ze^{2}q\cdot\frac{1}{x^{3}}(1-e^{-x}-xe^{-x})\;\;\;,\label{8}
\end{equation}
 to each electron state, where $v=4\pi R^{3}/3$ and $x=qR$; the
total change in energy $\Delta E$ is obtained by multiplying the
above result by the total number of electrons in the volume $v$,
which is given by $\int_{v}d{\bf r}\cdot n$; (the error made by counting
twice the interacting part of this energy (Koopmans' factor $1/2$)
is $Z^{2}e^{2}q(1+e^{-2x}-2e^{-x})/4\simeq0.07Z^{2}e^{2}q$, and it
can be neglected at this level of accuracy); one obtains 
\begin{equation}
\Delta E=-3Z^{2}e^{2}q\cdot\frac{1}{x^{3}}(1-e^{-x}-xe^{-x})^{2}=\frac{16}{3}E\cdot\frac{1}{x^{3}}(1-e^{-x}-xe^{-x})^{2}\;\;;\label{9}
\end{equation}
 this is a contribution to the total energy of the electrons, and
it must be minimized with respect to the parameter $R$, or, equivalently,
$x$, as noted above; the function of $x$ in equation (\ref{9})
has a maximum value $0.073$ for $x\simeq0.75$, which corresponds
to $R\simeq Z^{-1/3}a_{H}$, \textit{i.e.} close to the electronic
''radius'', and yields 
\begin{equation}
\Delta E=0.39E=-4.44Z^{7/3}{\rm eV}\;\;;\label{10}
\end{equation}
 therefore, the total energy is obtained as 
\begin{equation}
E=-11.4Z^{7/3}{\rm eV}-4.44Z^{7/3}{\rm eV}=-15.84Z^{7/3}{\rm eV}\;\;\;,\label{11}
\end{equation}
 which agrees well with the ''empirical'' binding energy $E\simeq-16Z^{7/3}{\rm eV}$.
Since the values derived here for the variational and the electronic
''radii'' are close to each other one may say that the computations
are consistent; one can see also that $\Delta E$ amounts to cca $28\%$
of the binding energy $E$, so that one may indeed regard $\Delta E$
as a correction to this energy; higher-order perturbation theory calculations
modify the electronic (quasi-) plane waves, and the single-particle
energies, according to the quantum behaviour; however, according to
the perturbation theory, the main contribution to the total energy
given above is not affected significantly. It is worth noting that
the quantum correction given above vanishes in the limit $Z\rightarrow\infty$,
as the electrons approach the quasi-classical limit; in addition,
the main contribution $-11.4Z^{7/3}{\rm eV}$ to the total energy
derived above is in error in the limit $Z\rightarrow\infty$, as the
linearization procedure is not valid anymore in this limit; indeed,
the linearization holds as long as the Fermi wavevector $k_{F}$ varies
slowly in space; a measure of the departure from this behaviour is
given by the extent to which the variational parameter $q=0.85Z^{1/3}/a_{H}$
given by equation (\ref{7}) differs from the parameter $q=(8Z/\pi^{2})^{1/3}/a_{H}\equiv0.9Z^{1/3}/a_{H}$
obtained from $q^{2}=4me^{2}\overline{k}_{F}/\pi\hbar^{2}=4\overline{k}_{F}/\pi a_{H}$,
where the average $\overline{k}_{F}$ is computed by using the electron
density $n({\bf r})=(q^{2}/4\pi e)\varphi(r)=(q^{2}Z/4\pi)e^{-qr}/r$
derived here; as one can see, the difference in the $q$-values is
$\sim5\%$, which implies a similar decrease in the total energy from
$-15.84Z^{7/3}{\rm eV}$ to $-16.64Z^{7/3}{\rm eV}$; this value has
the tendency to be more negative (towards the exact value $-20.8Z^{7/3}{\rm eV}$)
for large $Z$, as it ought to be. For finite values of $Z$ the error
in energy produced by the linearization procedure is nearly compensated
by the variational treatment of the quantum correction $\Delta E$.
This may explain the rather surprising proximity of the energy $E$
given by equation (\ref{11}) to the experimental atomic binding energy.
In this regard, one may say that the present linearized Thomas-Fermi
approach is more appropriate for an intermediate range of $Z$-values,
as corresponding to the ''actual'' atoms. The exchange energy must
be added to the result given above by equation (\ref{11}), and one
can check that it brings a $\sim4\%$ -correction at most, for $Z=20$.
As it can be seen easily, the $\Delta E$-correction computed here
in equation (\ref{10}) corresponds to the Hartree contribution to
the linearized Thomas-Fermi model; a similar correction to the exchange
energy can also be obtained; though very small, we give it here since
such corrections have previously been discussed to a rather large
extent, in the framework of the atomic theory;\cite{key-11,key-20}
on the other hand, the computation of such exchange corrections helps
to further enlighten the virtues of the linearized Thomas-Fermi model.

\noindent \textbf{Exchange energy.} As it is well known the exchange
energy of a homogeneous gas of electrons is given by $E_{ex}=-(e^{2}/4\pi^{3})Vk_{F}^{4}$;
according to the linearization procedure this energy is written as
\begin{equation}
E_{ex}=-\frac{e^{2}}{\pi^{3}}\overline{k}_{F}^{3}\int d{\bf r}\cdot k_{F}\,\,\,,\label{12}
\end{equation}
 and making use of the results obtained above, in particular $n=\overline{k}_{F}^{2}k_{F}/\pi^{2}$
and $n=(q^{2}/4\pi e)\varphi$ and the variational parameter $q$
derived in equation (\ref{7}), one obtains $E_{ex}=-18.12Z^{5/3}{\rm eV}$.
The correction to this exchange energy originates in the abrupt variation
of the electronic density near the atomic nucleus; to the first-order
of the perturbation theory it may be written as 
\begin{equation}
\Delta E_{ex}=-\frac{e^{2}}{(2\pi)^{6}}\int_{v}d{\bf r}\int d{\bf r}^{\prime}\int_{F}d{\bf k}d{\bf k}^{\prime}\cdot e^{-i{\bf Q\rho}}\frac{1}{\rho}\,\,\,,\label{13}
\end{equation}
 where ${\bf Q}={\bf k}-{\bf k}^{\prime}$, ${\bf \rho}={\bf r}-{\bf r}^{\prime}$,
$v$ is the spherical volume of radius $R$ around the nucleus, and
$F$ denotes the Fermi sea; in contrast to the Hartree correction
given by eqution (\ref{9}), the integration over ${\bf r}^{\prime}$
is extended over the whole space, as a consequence of the non-local
character of the exhange energy. The calculations in equation (\ref{13})
proceeds in the usual manner; first, we pass from the integration
over ${\bf r}^{\prime}$ to the integration over ${\bf \rho}$; the
result of this integration is $4\pi/Q^{2}-(4\pi/3)r(3R+2r)+...$;
one may neglect the small contribution of the second term, and retain
the main term $4\pi/Q^{2}$; next, we perform the ${\bf k},{\bf k}^{\prime}$-integrations,
which lead to 
\begin{equation}
\Delta E_{ex}=-\frac{e^{2}}{4\pi^{3}}\int_{v}d{\bf r}\cdot k_{F}^{4}\,\,;\label{14}
\end{equation}
 according to the linearization procedure equation (\ref{14}) may
also be written as 
\begin{equation}
\Delta E_{ex}=-\frac{e^{2}}{\pi^{3}}\overline{k}_{F}^{3}\int_{v}d{\bf r}\cdot k_{F}\;\;;\label{15}
\end{equation}
 one gets straightforwardly $\Delta E_{ex}=(1-e^{-x}-xe^{-x})E_{ex}$,
where $E_{ex}$ is given by equation (\ref{12}) and $x=qR$; for
the electronic ''radius'' $x\simeq1$ one obtains $\Delta E_{ex}\simeq0.27E_{ex}$,
while for the variational ''radius'' $x\simeq0.85$ derived above
one obtains $\Delta E_{ex}\simeq0.21E_{ex}$; it follows that the
exchange energy changes by a factor which lies somewhere between $1.21$
and $1.27$; it agrees well with similar exchange corrections derived
in Ref. \cite{key-11} (which indicates a factor $1.22$). It is customary
to refer such a factor in the exchange energy, denoted by $\alpha$,
to the value $2/3$, which corresponds to the homogenous electron
gas, \textit{i.e.} to $E_{ex}$ in the present calculations (and which
is known as the Kohn-Sham value\cite{key-21,key-22}); this is the
$\alpha$-factor in Slater's $X\alpha$-method (and in density-functional
calculations);\cite{key-20} according to the present results the
value of the $\alpha$-factor runs between $\alpha\simeq(2/3)\cdot1.21\simeq0.8$
and $\alpha\simeq(2/3)\cdot1.27\simeq0.85$; more accurate density-functional
computations\cite{key-20},\cite{key-23,key-24} of atomic and molecular
orbitals recommend $\alpha\simeq0.69-0.75$, which are in good agreement
with the present results (the terms neglected in the above ${\bf \rho}$-integration
diminish to some extent the value of the $\alpha$-factor); while
Slater's original value\cite{key-25} is $\alpha=1$.

\textbf{Giant dipole oscillations.} The electrons may move as a whole
with respect to the nucleus under the action of an external electric
field; for an oscillating external field the electrons may perform
giant dipole oscillations. During such small oscillations the equilibrium
is preserved, such as $Rq=const$; consequently, the small displacement
$u=\delta R$ is related to the change $\delta q$ in the screening
wavevector $q$ by $u=(1/q^{2})\delta q$. It follows that the change
in the energy arises solely from the change in the kinetic energy,
given in equation (\ref{5}); we get 
\begin{equation}
\delta E=\delta E_{kin}=\frac{27}{4\pi^{2}}\frac{Z^{3}e^{2}}{a_{H}^{3}}u^{2}\,\,;\label{16}
\end{equation}
 this energy corresponds to a frequency $\omega_{0}$ given by $\delta E=\frac{1}{2}M\omega_{0}^{2}u^{2}$,
where $M=Zm$ is the mass of all the electrons; we get the frequency
\begin{equation}
\omega_{0}=\left(\frac{27}{2\pi^{2}}\right)^{1/2}\frac{Ze}{\sqrt{ma_{H}^{3}}}\simeq4.5Z\times10^{16}s^{-1}\,\,;\label{17}
\end{equation}
 it corresponds to an energy $\hbar\omega_{0}\simeq28ZeV$, which
is in the range of moderate $X$-rays. The wavelength $\lambda_{0}=2\pi c/\omega_{0}\simeq\frac{4.2}{Z}\times10^{-6}cm$
is still much longer than the dimension of the atom ($c=3\times10^{10}$cm/s
is the speed of light). This is consistent with our adiabatic hypothesis
that during oscillations the equilibrium is preserved ($e=4.8\times10^{-10}statcoulomb$,
$m=10^{-27}g$, $\hbar=10^{-27}erg\cdot s$). 

As it is well known, an oscillating dipole radiates energy; consequently,
a damping force acts upon the dipole, given by $F_{d}=2Q^{2}\ddot{v}/3c^{3}$,
where $Q=-Ze$ is the charge of all the electrons and $v$ is the
velocity of the dipole; for the external frequency $\omega$ close
to the eigenfrequency $\omega_{0}$, we may put $\ddot{v}=\omega_{0}^{2}v$
and write 
\begin{equation}
F_{d}=\frac{2Z^{2}e^{2}}{3c^{3}}\omega_{0}^{2}v=M\gamma\dot{u}\,\,\,,\label{18}
\end{equation}
 where 
\begin{equation}
\gamma=\frac{2Ze^{2}}{3mc^{3}}\omega_{0}^{2}\label{19}
\end{equation}
 is a damping coefficient. We note that $\gamma\ll\omega_{0}$, since
$2Ze^{2}/3mc^{2}\ll c/\omega_{0}$, where $e^{2}/mc^{2}=r_{0}\simeq2.8\times10^{-13}cm$
is the classical electromagnetic radius of the electron ($c=3\times10^{10}cm/s$).
The quality ratio (natural breadth of the spectroscopic line) is 
\begin{equation}
\frac{\gamma}{\omega_{0}}=\frac{4\pi Zr_{0}}{3\lambda_{0}}\simeq2.8Z^{2}\times10^{-7}\,\,.\label{20}
\end{equation}
 Putting togeter all this information we can write the equation of
motion for the electrons 
\begin{equation}
M\ddot{u}+M\omega_{0}^{2}u+M\gamma\dot{u}=QE\cos\omega t\,\,\,,\label{21}
\end{equation}
 or 
\begin{equation}
m\ddot{u}+m\omega_{0}^{2}u+m\gamma\dot{u}=-eE\cos\omega t\,\,\,,\label{22}
\end{equation}
 where $E$ is the external electric field. We do not include the
effects of the magnetic field since the ratio $v/c$ is very small,
as we can see easily by comparing the kinetic energy with $Mc^{2}$.
Since the wavelength is much longer than the dimension of the atom
(quasi-stationary regime) we should include the effect of the internal
(polarization) field; we do not, since this field is valid only for
macroscopic bodies (or bodies with a definite surface). The (particular)
solution of equation (\ref{22}) is 
\begin{equation}
u=a\cos\omega t+b\sin\omega t\,\,\,,\label{23}
\end{equation}
 where 
\begin{equation}
a=\frac{eE}{m}\frac{\omega^{2}-\omega_{0}^{2}}{(\omega^{2}-\omega_{0}^{2})^{2}+\omega^{2}\gamma^{2}}\,,\,\, b=-\frac{eE}{m}\frac{\omega\gamma}{(\omega^{2}-\omega_{0}^{2})^{2}+\omega^{2}\gamma^{2}}\,\,.\label{24}
\end{equation}

We can see that the electrons perform giant dipole oscillations with
characteristic frequency (eigenfrequency) $\omega_{0}$. From the
energy conservation in equation (\ref{22}) we get the power loss
\begin{equation}
P=m\gamma\overline{\dot{u}^{2}}=\frac{e^{2}E^{2}}{2m}\frac{\gamma\omega^{2}}{(\omega^{2}-\omega_{0}^{2})^{2}+\omega^{2}\gamma^{2}}\,\,\,,\label{25}
\end{equation}
 which, at resonance, becomes $P_{res}=e^{2}E^{2}/2m\gamma$. Making
use of equations (\ref{17}) and (\ref{20}), for moderate fields
$E=1/300statvolt/cm$ ($100V/m$) we get a power loss 
\begin{equation}
P_{res}=\frac{1}{Z^{3}}\times10^{-7}erg/s\,\,;\label{26}
\end{equation}
 it corresponds to a transition rate 
\begin{equation}
R=P_{res}/\hbar\omega_{0}\simeq\frac{2}{Z^{4}}\times10^{3}s^{-1}\,\,\,,\label{27}
\end{equation}
where $R$ represents the number of elementary acts of oscillation
per unit time. The formula given by equation (\ref{25}) is valid
for one electron; for the atom we must multiply equation (\ref{25})
by $Z$. The transition rate remains unchanged, because each electron
is an oscillator which absorbs (and emits) an energy quanta $\hbar\omega$
($\hbar\omega_{0})$. 

The linearized Thomas-Fermi model allows the estimation of a motion
involving only $1\ll\delta Z<Z$ electrons, the rest of $Z-\delta Z$
electrons together with the atomic nucleus being considered as an
inert core with charge $\delta Z\cdot e$. The characteristic frequency
given by equation (\ref{17}) becomes $(\delta Z/Z)\omega_{0}$ and
the damping coefficient is $(2e^{2}\delta Z/3mc^{3})\omega_{0}^{'2}$. 

\textbf{Anharmonicities and ionization.} It is worth estimating the
amplitude of oscillations, which, at resonance, is given by 
\begin{equation}
\left|b_{0}\right|=\frac{eE}{m\omega_{0}\gamma}=\frac{8}{Z^{4}}\times10^{-10}E\, cm\,\,\,,\label{28}
\end{equation}
 from equations (\ref{17}), (\ref{20}) and (\ref{23}). For an extended
range of field intensities this amplitude is much smaller than the
characteristic distances in atom, for instance, the Bohr radius. Consequently,
the harmonic approximation is justified. Higher-order corrections
to the harmonic approximation are obtained by including higher-order
terms in the kinetic energy given in equation (\ref{5}) and using
the relationship $(q+\delta q)(R+u)=1$ ($\delta q=-q^{2}u/(1++qu)$,
which corresponds to the adiabatic perturbation, valid as long as
the oscillation frequency is much smaller than the frequency of the
quantum one-particle state). We get a potential energy 
\begin{equation}
U=\frac{1}{2}M\omega_{0}^{2}u^{2}\left(1-\frac{8}{3}qu+\frac{31}{6}q^{2}u^{2}+...\right)\,\,,\,\, q=(6/\pi^{2})^{1/3}Z^{1/3}/a_{H}=0.85Z^{1/3}/a_{H}\,\,\,,\label{29}
\end{equation}
which contains anharmonic terms. As it is well known, the corresponding
non-linear (free) oscillations imply higher-order harmonics (with
frequencies $2\omega_{0}$, $3\omega_{0}$, etc), displacements of
the equilibrium position and shifts in the original frequency $\omega_{0}$,
which may be computed either by successive approximations or by the
self-consistent harmonic-oscillator approximation. In particular,
the frequency shifts are worth noting, since they determine abrupt
changes in the oscillation amplitude near resonance ($\omega\simeq\omega_{0}$);
due to the combined frequencies phenomenon, the $\omega_{0}$-resonance
may be excited by other frequencies, or other resonances may be excited.\cite{key-26} 

Since $\delta q=-q^{2}u/(1+qu)$, for large oscillation amplitudes
($u\rightarrow\infty$) we get the energy $U_{\infty}=9Z^{2}e^{2}q/16$
(with $q=0.85Z^{1/3}/a_{H}$); this energy cancels out exactly the
binding energy given by equation (\ref{6}), as expected, setting
the electrons free (the quantum energy correction is vanishing in
this limit). Therefore, in order to have complete ionization (hyper-ionization,
dissociation) we should compare the amplitude given by equation (\ref{28})
with the Bohr radius; we get 
\begin{equation}
E>7Z^{4}\,\,;\label{30}
\end{equation}
 this is a high field, for ($X$-ray) frequency $\omega_{0}$. (High-intensity
fields are generated in short laser pulses; for instance, for intensity
$I=10^{15}w/cm^{2}$ we get an electric field $E=10^{6}statvolt/cm$
($E\simeq\sqrt{I/c)}$); this is an \textquotedbl{}atomic field\textquotedbl{},
of the order of the electron field in atoms; it may generate ionization
and high-order harmonics. For intensities $10^{19}w/cm^{2}$ (curent
intensities) the field is of the order $10^{8}statvolt/cm$, where
relativistic and non-linear effects apear; for ultrahigh intensities
$10^{21}w/cm^{2}$ the field is $10^{9}statvolt/cm$, where multi-photon
processes appear, the structure of the quantum vacuum may also occur,
as well as particle production (the Schwinger field, which indicates
a limit of quantum electrodyamics calculations, is $10^{13}statvolt/cm$).
All these fields are optical fields (energy $\simeq1eV$, frequency
$\simeq10^{15}s^{-1}$, wavelength $\simeq10^{-4}cm=1\mu m$); typically,
the pulse duration is $50fs$ ($1fs=10^{-15}s$) and the pulse dimension
is $d\simeq15\mu m$; for intensity $10^{21}w/cm^{2}$ the power is
$P\simeq1pw$ ($1p=10^{15}$).). For the motion of a fraction $\delta Z$
of electrons ($1\ll\delta Z<Z$) we have $\omega_{0}=4.5\delta Z\times10^{16}s^{-1}$
and $\gamma/\omega_{0}=2.8(\delta Z)^{2}\times10^{-7}$. We can see
that $\omega_{0}$ may lie now in the ultraviolet range. $\hbar\omega_{0}$
can be viewed as an ionization energy; for $\delta Z=1$ we get $\hbar\omega_{0}=30eV$,
which is higher than the (first) ionization potential of the elements
(an average of $6eV$); we note that the extrapolation to $\delta Z=1$
is not permissible. The critical electric field for partial ionization
is given by 
\begin{equation}
E>7(\delta Z)^{4}\,\,\,,\label{31}
\end{equation}
 which is much smaller than the field for total ionization given by
equation (\ref{30}). 

\textbf{Quasi-classical approximation.} As it is well known, the equation
of motion for an operator $O$ reads $\dot{O}=\frac{i}{\hbar}[H,O]$,
or $\dot{O}_{mn}=\frac{i}{\hbar}(E_{m}-E_{n})O_{mn}=i\omega_{mn}O_{mn}$,
where $H$ is the hamiltonian, $E_{n}$, $E_{m}$ are the energies
of the states $n$ and, respectively, $m$ and $\omega_{mn}=(E_{m}-E_{n})/\hbar$
is the frequency of transition between the states $n$ and $m$. In
the quasi-classical approximation the quantum states are sufficiently
dense to approximate the frequency $\omega_{mn}$ by $\omega_{mn}\simeq-s(\partial E_{m}/\partial m)_{m}=-\omega_{s}$,
where $n=m+s$ and $E_{n}$ depends slightly on $n$ in the vicinity
of $m$; this amounts to a quasi-classical motion which implies a
mechanical action much greater than $\hbar$. Similarly, for a set
of quantum states sufficiently dense the matrix elements $O_{mn}=O_{m,m+s}$
depends slightly on $s$ for small $s$, and vanishes rapidly for
greater $s$, so that we may write $O_{mn}=O_{m,m+s}\simeq O_{s}$;
in fact, $O_{s}$ is the temporal Fourier transform of $O$, corresponding
to the frequency $\omega_{s}$; (the average of $O$ with the wavefunction
$\psi=\sum_{n}c_{n}\varphi_{n}e^{-i\omega_{n}t}$ is $\overline{O}=\sum_{mn}c_{m}^{*}c_{n}O_{mn}e^{i\omega_{mn}t}=\sum_{ms}c_{m}^{*}c_{m+s}O_{m,m+s}e^{-i\omega_{s}t}$,
which is approximateley $\overline{O}\simeq\sum_{m}\left|c_{m}\right|^{2}\sum_{s}O_{s}e^{-i\omega_{s}t}\simeq\sum_{s}O_{s}e^{-i\omega_{s}t}$;
$O_{s}(t)=O_{s}e^{-i\omega_{s}t}$ is the time-dependent operator
in the quasi-classical equation of motion (\ref{32})); we may drop
out the label $s$ of the $O_{s}$, and we may add an external force,
as represented by a hamiltonian $h$, in general time-dependemt; the
equation of motion becomes $\dot{O}=-i\omega_{s}O+(\partial O^{cl}/\partial t)_{cl;h}$,
where the last term means the time derivative of the classical counterpart
$O^{cl}$ of $O$, as given by $h$, according to the classical motion.
With $O=O^{(1)}+iO^{(2)}$, we get $\dot{O}^{(1)}=\omega_{s}O^{(2)}+(\partial O^{cl}/\partial t)_{cl;h}$,
$\dot{O}^{(2)}=-\omega_{s}O^{(1)}$ (since the classical quantity
$(\partial O^{cl}/\partial t)_{cl;h}$ is real); we get $\ddot{O}^{(1)}+\omega_{s}^{2}O^{(1)}=(\partial/\partial t)(\partial O^{cl}/\partial t)_{cl;h}$;
here, we may identify $O^{(1)}$ with the time-dependent part of the
classical quantity $O^{cl}$, and leave aside the labels (similarly
for $O^{(2)}$); (the general solution for $O^{(1)}$ from the homogeneous
version of equation (\ref{32}) is $O^{(1)}=A\cos(\omega_{s}t+\delta)$,
where $A$ is amplitude and $\delta$ is a phase, both undetermined;
from $\dot{O}^{(2)}=-\omega_{s}O^{(1)}$, we get $O^{(2)}=-A\sin(\omega_{s}t+\delta)$,
and $O=O^{(1)}+iO^{(2)}=Ae^{-i(\omega_{s}t+\delta)}$, as expected;
the latter ($\sim e^{-i\omega_{s}t}$, or $e^{i\omega_{s}t}$) is
the quantum version (in the quasi-classical approximation), while
the fomer ($\sim\cos\omega_{s}t$, or $\sin\omega_{s}t$) is the classical
version of the same quantity. ); we get 
\begin{equation}
\ddot{O}+\omega_{s}^{2}O=(\partial/\partial t)(\partial O^{cl}/\partial t)_{cl;h}=f\,\,;\label{32}
\end{equation}
 this is the equation of motion of a classical harmonic oscillator
subjected to the action of a classical force $f$; (the force $f$
should contain only $c$-numbers (if necessary, they can be determined
by comparing the absorbed power computed both classically and quantum-mechanically).
); its eigenfrequency $\omega_{s}$ is the quantum transition frequency
$\omega_{mn}$ in the quasi-classical approximation. Since the approximation
is valid for a wavepacket, we may also introduce a lifetime $\gamma^{-1}$
given by a damping term $\gamma\dot{O}$ in equation (\ref{32}). 

The \textquotedbl{}peripheral\textquotedbl{} electrons in atom, \emph{i.e.}
the electrons with high quantum numbers (and high energy) in the mean-field
atomic potential, which may be subjected to ionization by the action
of an (optical) electromagnetic field, satisfy the conditions for
the quasi-classical approximation in heavy atoms; under the action
of the hamiltonian $h=eEu\cos\omega t$, force $f=-eE\cos\omega t$,
where $E$ is the electric field and $u$ is the displacement, the
equation of motion of such an electron is 
\begin{equation}
\ddot{u}+\omega_{0}^{2}u+\gamma\dot{u}=-\frac{e}{m}E\cos\omega t\,\,\,,\label{33}
\end{equation}
 where $\omega_{0}$ ($=\omega_{s}$) is the ionization frequency
(excited states are described similarly). The amplitude at resonance
is given by $\left|b_{0}\right|=eE/m\omega_{0}\gamma$ (equation (\ref{28})),
where the damping coefficient is $\gamma=2r_{0}\omega_{0}^{2}/3c=6\times10^{-24}\omega_{0}^{2}$
(equation (\ref{19})). We take $\omega_{0}=10^{16}s^{-1}$ ($\hbar\omega_{0}=6eV$,
average ionization potential) and get $\left|b_{0}\right|=10^{-7}E\, cm$;
compared with the Bohr radius, it leads to a critical field of ionization
$E>5\times10^{-2}statvolt/cm$ ($\simeq1500V/m$). For higher fields
we may have multiple-quanta transitions; they correspond to larger
displacements for an oscillator, when the harmonic approximation does
not hold anymore. The mean-field potential $U$ which gives $\omega_{0}^{2}=(1/m)(\partial^{2}U/\partial u^{2})_{0}$
may contribute now higher-order terms like $\sim u^{3},$$\sim u^{4}$,
which leads to anharmonicities in the classical equation of motion
(\ref{33}). 

\textbf{Conclusion.} In conclusion, one may say that the variational
treatment of the linearized Thomas-Fermi model provides a consistent
quasi-classical description for the atomic binding energies in the
range of realistic values of atomic numbers $Z$ (heavy atoms), provided
the quantum corrections (Hartree-type contributions) are properly
included. (This might be expected since the ''boundary effect''
included in the asymptotic series originates in the quantum corrections
too (see, for instance, Refs. \cite{key-9} and \cite{key-10}, and
discussion therein)). Making use of the Thomas-Fermi model it was
shown here that giant dipole oscillations may be induced in heavy
atoms by external electromagnetic fields in the moderate $X$-ray
range, which may lead to the ionization in intense fields. Frequency
shifts and higher-order harmonics can be produced by anharmonicities
in the dipole oscillations. Quasi-classical equation of motion for
\textquotedbl{}peripheral\textquotedbl{} electrons was also derived
(a harmonic-oscillator equation), which can be used for investigating
transitions to excited states or ionization in heavy atoms. 

\textbf{Appendix}

\textbf{Exchange energy}

The exchange energy of a set of electrons which interact through the
Coulomb potential (the Fock term) is given by 
\begin{equation}
E_{ex}=-\frac{e^{2}}{2V^{2}}\int d\mathbf{r}d\mathbf{r}^{'}\sum_{\mathbf{k}\mathbf{k}^{'}(\uparrow\uparrow)}\frac{\varphi_{\mathbf{k}^{'}}^{*}(\mathbf{r}^{'})\varphi_{\mathbf{k}}(\mathbf{r}^{'})}{\left|\mathbf{r}-\mathbf{r}^{'}\right|}\varphi_{\mathbf{k}}^{*}(\mathbf{r})\varphi_{\mathbf{k}^{'}}(\mathbf{r})\,\,\,,\label{34}
\end{equation}
 where $-e$ is the electron charge, $V$ is the volume enclosing
the set of electrons, $\varphi_{\mathbf{k}}(\mathbf{r})$ are the
single-particle wavefunctions and the summation includes parallel
spins; the summation over $\mathbf{k}$, $\mathbf{k}^{'}$ is performed
over the Ferm sea. For plane waves $\varphi_{\mathbf{k}}(\mathbf{r})=e^{i\mathbf{kr}}$
we get 
\begin{equation}
E_{ex}=-\frac{e^{2}}{V}\sum_{\mathbf{k}\mathbf{k}^{'}}\int d\mathbf{r}\frac{e^{i\mathbf{qr}}}{r}\,\,\,,\label{35}
\end{equation}
 where $\mathbf{q}=\mathbf{k}-\mathbf{k}^{'}$. The Fourier transform
of the Coulomb potential is $\int d\mathbf{r}e^{i\mathbf{qr}}/r=4\pi/q^{2}$,
so the exchange energy becomes
\begin{equation}
E_{ex}=-\frac{2e^{2}}{(2\pi)^{5}}V\int d\mathbf{k}d\mathbf{k}^{'}\frac{1}{q^{2}}\,\,.\label{36}
\end{equation}
We introduce the new variables $\mathbf{q}=\mathbf{k}-\mathbf{k}^{'}$
and $\mathbf{p}=\mathbf{k}+\mathbf{k}^{'})/2$; it is easy to see
that the integration over $\mathbf{p}$ extends over the intersection
of two Fermi see separated by $\mathbf{q}$, $0<q<2k_{F}$, where
$k_{F}$ is the Fremi wavevector. This intersection consists of two
equal spherical sectors, subtended by the angle $\theta_{0}$ given
by $\cos\theta_{0}=q/2k_{F}$. The volume of a spherical sector is
\begin{equation}
v=\int\pi k_{F}^{2}\sin^{2}\theta\cdot k_{F}d\theta\sin\theta=\frac{2\pi}{3}k_{F}^{3}(1-\frac{3}{2}\cos\theta_{0}+\frac{1}{2}\cos^{3}\theta_{0})\,\,.\label{37}
\end{equation}
 We get the exchange energy 
\begin{equation}
E_{ex}=-\frac{8e^{2}}{3(2\pi)^{3}}Vk_{F}^{3}\int_{0}^{2k_{F}}dq\left(1-\frac{3}{2}\frac{q}{2k_{F}}+\frac{1}{2}\frac{q^{3}}{(2k_{F})^{3}}\right)=-\frac{e^{2}}{4\pi^{3}}Vk_{F}^{4}\,\,.\label{38}
\end{equation}
 It is usual to introduce the inter-particle separation $r_{s}$ through
$n=k_{F}^{3}/3\pi^{2}=1/(4\pi r_{s}^{3}/3)$, \emph{i.e.} $k_{F}r_{s}=(9\pi/4)^{1/3}$;
the exchange energy per electron is written as 
\begin{equation}
E_{ex}/N=-\frac{e^{2}}{2a_{H}}\frac{2}{3\pi^{2}}\left(\frac{9\pi}{4}\right)^{4/3}\frac{1}{r_{s}}=-\frac{0.916}{r_{s}}ry\,\,\,,\label{39}
\end{equation}
 where $N$ is the number of electrons, $r_{s}$ is measured in Bohr
radii ($a_{H}=\hbar^{2}/me^{2}=0.53\textrm{\AA}$) and the energy
is measured in rydbergs ($1ry=e^{2}/2a_{H}=13.6eV$). Similarly, the
kinetic energy leads to $E_{kin}/N=2.21/r_{s}^{2}ry$ (next-order
contributions to the perturbation series, which give the correlation
energy, are $0.062\ln r_{s}-0.094$).\cite{key-27} 

\textbf{Koopmans' factor.} We may have an estimation of the effect
of the Koopman's factor $1/2$ by comparing the energy correction
\begin{equation}
\Delta E=\int_{v}d\mathbf{r}\rho_{e}\varphi=-Z^{2}e^{2}q^{2}\int_{0}^{R}dre^{-2qr}=-\frac{1}{2}Z^{2}e^{2}q(1-e^{-2qR})\label{40}
\end{equation}
 with 
\begin{equation}
\Delta E^{'}=\int_{v}d\mathbf{r}(\rho_{e}\varphi-\frac{1}{2}\rho_{e}\varphi_{e})=\frac{1}{2}\int_{v}d\mathbf{r}\rho_{e}(\varphi+\varphi_{c})\,\,\,,\label{41}
\end{equation}
 where $\rho_{e}=-en$ is the electron charge density, $n$ is the
electron density, $\varphi_{e}=\varphi-\varphi_{c}$ is the electron
potential and $\varphi_{c}=Ze/r$ is the potential of the nucleus
(core potential). We get 
\begin{equation}
\Delta E^{'}=-\frac{1}{4}Z^{2}e^{2}q(3-e^{-2qR}-2e^{-qR})\label{42}
\end{equation}
 and 
\begin{equation}
\Delta E-\Delta E^{'}=\frac{1}{4}Z^{2}e^{2}q(1+e^{-2qR}-2e^{-qR})\,\,\,,\label{43}
\end{equation}
 which is indeed very small ($\simeq0.07Z^{2}e^{2}q$ ) for $qR=0.75$. 

In fact, Koopmans' factor does not appear in the energy correction
given by equations (\ref{8}) and (\ref{9}) since the potential $\varphi$
is mainly determined by electrons lying away from the nucleus, while
the quantum correction implies electrons placed close to the nucleus;
or, in other words, the quantum correction, which implies the strong
variation of the potential $\varphi$, \emph{i.e.} an appreciable
deviation from the quasi-classical approximation, is deteAZrmined
mainly by the nucleus. 

\textbf{Quantum and classical transitions.} The quantum transition
amplitude from state $n$ to state $k$, energies $E_{k}$ and, respectively,
$E_{n}$, under the action of a perturbation $V(t)=V\cos\omega t$,
in the first order of the perturbation theory, is given by 
\begin{equation}
i\hbar\dot{c}_{k}=V_{kn}(t)e^{i\omega_{kn}t}\,\,\,,\label{44}
\end{equation}
 or 
\begin{equation}
c_{kn}=-\frac{i}{2\hbar}\int_{-\infty}^{t}dt^{'}V_{kn}\left[e^{i(\omega+\omega_{kn})t+\alpha t}+e^{i(-\omega+\omega_{kn})t+\alpha t}\right]\,\,\,,\label{45}
\end{equation}
where $\omega_{kn}=(E_{k}-E_{n})/\hbar$the interaction is introduced
adiabaticaly ($\alpha\rightarrow0^{+}$); we retain only transitions
with $\omega>0$, and get 
\begin{equation}
c_{k}=\frac{V_{kn}}{2\hbar}\frac{e^{i(-\omega+\omega_{kn})t+\alpha t}}{\omega-\omega_{kn}+i\alpha}\,\,.\label{46}
\end{equation}
The transition rate (number of transitions per unit time) is given
by 
\begin{equation}
R=\frac{\partial\left|c_{k}\right|^{2}}{\partial t}=\frac{\left|V_{kn}\right|^{2}}{2\hbar^{2}}\frac{\alpha}{(\omega-\omega_{kn})^{2}+\alpha^{2}}\rightarrow\frac{\pi\left|V_{kn}\right|^{2}}{2\hbar^{2}}\delta(\omega-\omega_{kn})\,\,.\label{47}
\end{equation}
 For a dipolar interaction $V=eEu$ ($-\mathbf{dE}$, where $d$ is
the electric dipole moment) we get 
\begin{equation}
R=\frac{e^{2}E^{2}\left|u_{kn}\right|^{2}}{2\hbar^{2}}\frac{\alpha}{(\omega-\omega_{0})^{2}+\alpha^{2}}\,\,\,,\label{48}
\end{equation}
or, for $\omega_{kn}=\omega_{0}$ (\textquotedbl{}quantum\textquotedbl{}
oscillations), $k=1$, $n=0$, $u_{10}=\sqrt{\hbar/2m\omega_{0}}$,
\begin{equation}
R=\frac{e^{2}E^{2}}{4m\hbar\omega_{0}}\frac{\alpha}{(\omega-\omega_{0})^{2}+\alpha^{2}}\simeq\frac{e^{2}E^{2}}{2m\hbar\omega_{0}}\frac{2\alpha\omega^{2}}{(\omega^{2}-\omega_{0}^{2})^{2}+4\omega^{2}\alpha^{2}}\,\,\,,\label{49}
\end{equation}
 which coincides with equations (\ref{25}) and (\ref{27}) for classical
oscillations (\textquotedbl{}classical transitions\textquotedbl{})
for $\omega\simeq\omega_{0}$ and $2\alpha=\gamma$. 

It may appear disagreeable that the same result is obtained for \textquotedbl{}quantum
transitions\textquotedbl{}, \emph{i.e.} transitions from the ground-state
to the first excited state of the oscillator, and the \textquotedbl{}classical
transitions\textquotedbl{}, \emph{i.e.} oscillations of the classical
oscillator. The explanation resides in the fact that the classical
harmonic oscillator is restricted to small oscillations, which correspond
in fact to quantum motion, while the quantum oscillator allows also
large displacements, which may trespass the harmonic-oscillations
criterion of small displacement (the matrix elements of the displacement
$x$ for a harmonic oscillator with frequency $\omega$, mass $m$
are $x_{n,n-1}=\sqrt{n\hbar/2m\omega}$). In this context, the quasi-classical
approximation gives only the small variation of the quantum transitions.

\textbf{Dipolar radiation.} It is worth comparing the radiated intensity
to the power loss. The radiation intensity (energy per unit time)
is given by 
\begin{equation}
I=\frac{2e^{2}}{3c^{3}}\overline{\ddot{u}^{2}}=\frac{1}{3}E^{2}r_{0}^{2}c\frac{\omega^{4}}{(\omega^{2}-\omega_{0}^{2})^{2}+\omega^{2}\gamma^{2}}\label{50}
\end{equation}
 for a dipole $-eu$. Making use of the power loss given by equation
(\ref{25}), 
\begin{equation}
P=\frac{e^{2}E^{2}}{2m}\frac{\gamma\omega^{2}}{(\omega^{2}-\omega_{0}^{2})^{2}+\omega^{2}\gamma^{2}}\,\,\,,\label{51}
\end{equation}
 we get $I/P=2r_{0}\omega^{2}/3c\gamma$. For $\omega=\omega_{0}$,
making use of equation (\ref{19}) (with $Z=1$), we get $I=P$. The
result does not depend on $Z$, such that it holds also for atom. 

\textbf{Acknowledgments.} The first part of this article is based
on a previous publication with L. C. Cune (Roum. J. Phys. \textbf{55}
913-919 (2010)). The author is indebted to the members of the Laboratory
of Theoretical Physics at Magurele-Bucharest for useful discussions.
This work was supported UEFISCDI Grants Core Program \#09370108-2009/Ph12/2014
of the Romanian Governmental Agency for Research.

\end{document}